\documentclass[aps,prb,superscriptaddress,twocolumn,longbibliography]{revtex4-1}
\usepackage{bbm}
\usepackage{graphicx}
\usepackage{dcolumn}
\usepackage{bm}
\usepackage{amsmath}
\usepackage{feynmf}
\usepackage{hyperref}
\usepackage{subfigure}  
\usepackage{bm}
\usepackage{times}
\usepackage{color}
\usepackage{attachfile}

\newcommand{\bk}{\boldsymbol{k}}
\newcommand{\bq}{\boldsymbol{q}}

\newcommand{\ba}{\boldsymbol{a}}
\newcommand{\bb}{\boldsymbol{b}}
\newcommand{\bM}{\boldsymbol{M}}
\newcommand{\bs}{\boldsymbol{s}}
\newcommand{\Neel}{N\'{e}el }
\newcommand{\bd}{\boldsymbol{d}}

\newcommand{\zb}{\color {black}}

\begin{document}

\title{Intrinsic nonlinear Hall effect in two-dimensional honeycomb topological antiferromagnets}

\author{Zheng-Yang Zhuang}
\affiliation{Guangdong Provincial Key Laboratory of Magnetoelectric Physics and Devices,
State Key Laboratory of Optoelectronic Materials and Technologies,
and School of Physics, Sun Yat-sen University, Guangzhou 510275, China}

\author{Zhongbo Yan}
\email{yanzhb5@mail.sysu.edu.cn}
\affiliation{Guangdong Provincial Key Laboratory of Magnetoelectric Physics and Devices,
State Key Laboratory of Optoelectronic Materials and Technologies,
and School of Physics, Sun Yat-sen University, Guangzhou 510275, China}

\date{\today}

\begin{abstract}
Two-dimensional systems with honeycomb lattice are known to be a paradigmatic platform to explore
the various types of Hall effects, owing to that the interplay of lattice geometry,
spin-orbit coupling and magnetism can give rise to very rich features in the quantum
geometry of wave functions.
In this work, we consider honeycomb topological antiferromagets that are
effectively described by a $\mathcal{PT}$-symmetric antiferromagnetic Kane-Mele model,
and explore the evolution of its nonlinear Hall response with respect to the
change of lattice anisotropy, chemical potential, and the direction of the
\Neel vector. Due to the $\mathcal{PT}$-symmetry,
the leading-order Hall effect of quantum geometric origin is the {\zb time-reversal-odd} intrinsic nonlinear Hall effect, which is a second-order effect of electric
fields and is independent of the scattering time.  We investigate the behavior
of the intrinsic nonlinear Hall conductivity tensor across topological phase transitions driven by
antiferromagnetic exchange field and lattice anisotropy and find that its components
do not change sign, which is different from the {\zb time-reversal-even} nonlinear Hall effect {\zb of Berry curvature dipole origin}.
In the weakly doped regime, we find that the intrinsic nonlinear Hall effect
is valley-polarized. By varying the chemical potential, we find that the nonlinear Hall conductivity tensors
exhibit kinks when the Fermi surface undergoes Lifshitz transitions.
Furthermore, we find that the existence of spin-orbit coupling to lift the spin-rotation symmetry is decisive for
the use of intrinsic nonlinear Hall effect to detect the direction of the \Neel vector.
Our work shows that the two-dimensional honeycomb topological antiferromagnets are an ideal
 class of material systems with rich properties for the study of intrinsic nonlinear Hall effect.
\end{abstract}


\maketitle

\section{Introduction}

The quantum geometry of wave functions has
a fundamental and deep connection with the behavior of electrons. Two basic quantum geometric
quantities are the quantum metric and Berry curvature, which correspond to
the real part and imaginary part of the so-called quantum geometric tensor~\cite{Provost1980}, respectively.
The Berry curvature has been extensively studied over the past few decades and revealed
to be an indispensable factor to understand many important phenomena in materials, with the most
celebrated example being its application in understanding
the quantized (anomalous) Hall effect~\cite{thouless1982,Chang2023QAHE} and the generic non-quantized anomalous
Hall effect in magnetic metals~\cite{Nagaosa2010AHE} or topological semimetals~\cite{yang2011}. Compared to the Berry curvature,
the quantum metric started to attract wide interest in the condensed-matter field
much more lately. The reason is partly due to that the quantum metric influences the electrons relatively more
subtly, unlike the Berry curvature that gives a transparent contribution to the velocity operator~\cite{Xiao2010BP}.
Nevertheless, recent studies have shown that the quantum metric is also fundamentally important
for the understanding of many important phenomena, such as the superconductivity in
flat bands~\cite{Peotta2015,Julku2016,Liang2017geometry,Hu2019geometry,Xie2020geometry,Huhtinen2022},
optical responses~\cite{Chen2021geometry,Ahn2022,Onishi2023}, etc~\cite{Torma2023}.

In the past few years, the generalized higher-order moments of
Berry curvature and quantum metric have further generated considerable interest
as they can induce Hall-type effects in the nonlinear response regime~\cite{Moore2010BP,Gao2014INHE,Sodemann2015BCD,Xu2018BCD,Zhang2018BCDweyl,
You2018BCD,Zhang2018BCD,Du2018NHE,Facio2018,Ma2019NHE,Kang2019,Battilomo2019BCD,Wang2019NHE,Xiao2019NHE,
Rostami2020,Sinh2020NHE,Zeng2020BCD,Pantaleon2020BCD,
Satyam2020BCD,Kumar2021NHE,Liao2021NHE,Zhang2022NHE,
Bandyopadhyay2022NLHE,Roy2022NHLE,Okyay2022,Sinha2022,Chakraborty2022,Kamal2023metric,
Wang2023INHE,Zhang2023multipole,Atencia2023NLHE,Zhuang2023NLHE,
Huang2023INHE,Saha2023NLHE,Kirikoshi2023INAH,Balazs2023geometry,Huang2023NLHE,Kaplan2024NLHE,Mandal2023,Ma2021review,Du2021review,Ortix2021,Arka2024review}.
The two nonlinear Hall effects (NLHEs) that have attracted particular interest are the
{\zb time-reversal-even  NLHE of Berry-curvature-dipole origin~\cite{Sodemann2015BCD} (for the convenience of discussion,
we dub it as Berry-curvature-dipole NLHE) and the time-reversal-odd NLHE of quantum-metric-dipole (or say Berry connection 
polarizability) origin~\cite{Gao2014INHE} (known
as intrinsic NLHE) in inversion-asymmetric systems.  Because of the fundamental difference under time reversal,
the Berry-curvature-dipole NLHE can appear in a time-reversal invariant system, whereas
the intrinsic NLHE can only show up in systems without time-reversal symmetry.
The adjective ``intrinsic'' refers to the fact that the effect does not
depend on the scattering time and only depends on the band property.}
Being a time-reversal odd effect, the intrinsic NLHE has been shown in theory that
it holds promise for applications in antiferromagnetic spintronics as it
has the power to detect one key property of the antiferromagnets,
the \Neel vector~\cite{Liu2021INHE,Wang2021INHE}. Besides the prospect of applications in spintronics, the detection of
the direction of \Neel vector is also of significant importance in its own right, since
many properties of an antiferromagnet, such as band topology~\cite{Libor2017,Xu2019HOTI},
sensitively depend on it. Remarkably, the intrinsic NLHE and its sign change upon reversing the
\Neel vector have recently been experimentally observed in even-layered topological antiferromagnets~\cite{Gao2023metric,Wang2023metric}, MnBi$_{2}$Te$_{4}$.
This breakthrough has paved the way to explore the interplay of antiferromagnetism and
other factors of a system through the intrinsic NLHE in experiment.

The lattice structure, spin-orbit coupling and magnetism are
three factors that strongly influence the band structure
and the quantum geometry of the Bloch wave functions. When the band structure sensitively
depends on their interplay, it is natural to expect that the intrinsic NLHE
would exhibit characteristic features.
Among various lattice structures, the two-dimensional (2D) honeycomb lattice
is known to be a paradigmatic platform where the spin-orbit coupling and magnetism
can influence the band topology in a nontrivial way~\cite{Niu2017qshe,Zou2022AFMTI,Li2022AFMTI}. Therefore, a
honeycomb topological antiferromagnet is expected to be an ideal platform to explore the
intrinsic NLHE~\cite{Wang2023INHE}. With this picture in mind, in this work we consider
honeycomb topological antiferromagets effectively described
by a $\mathcal{PT}$-symmetric antiferromagnetic
Kane-Mele model and explore the evolution of the intrinsic
NLHE with respect to the change of lattice anisotropy, band topology,
chemical potential, and the direction of the \Neel vector. Our
main findings include: (i) the lattice anisotropy breaking the
$\mathcal{C}_{3z}$ rotation symmetry is crucial for having
a nonzero intrinsic NLHE; (ii) when the band topology
changes from a quantum spin Hall insulator to a trivial insulator or a boundary-obstructed atomic insulator,
the intrinsic NLHE preserves its direction, which is distinct from the {\zb Berry-curvature-dipole} NLHE; (iii) the intrinsic NLHE
is valley-polarized in the weakly-doped regime and exhibit nonanalyticity when the Fermi surface undergoes Lifshitz transitions;
(iv) the existence of spin-orbit coupling to lift the spin-rotation symmetry is decisive for
detecting the direction of the \Neel vector.
These results suggest that the intrinsic NLHE provides an effective tool to
measure basic properties of 2D honeycomb topological antiferromagnets.

The paper is organized as follows. In Sec.\ref{II}, we give the
effective tight-binding Hamiltonian and discuss the
important symmetries and possible topological phases associated
with the Hamiltonian. In Sec.\ref{III}, we investigate the behavior
of the intrinsic NLHE across two types of topological phase transitions.
In Sec.\ref{IV}, we study the dependence of the intrinsic NLHE on
the chemical potential and the direction of the \Neel vector. In
Sec.\ref{V}, we discuss our findings and conclude the paper.

\section{Theoretical model}\label{II}

A $\mathcal{PT}$-symmetric honeycomb collinear antiferromagnet with finite
intrinsic spin-orbit coupling can be effectively described by the tight-binding Kane-Mele model.
The Hamiltonian is given by $H=\sum_{\bk}\Psi_{\bk}^{\dagger}\mathcal{H}(\bk)\Psi_{\bk}$, where the basis is chosen as $\Psi_{\bk}^{\dagger}=(c_{A,\uparrow,\bk}^{\dagger},c_{B, \uparrow,\bk}^{\dagger},c_{A,\downarrow,\bk}^{\dagger},c_{B,\downarrow,\bk}^{\dagger})$
 and the momentum-space Hamiltonian reads~\cite{Kane2005a,Kane2005b}
\begin{eqnarray}
	\mathcal{H}(\bk)&=&\sum_{i=1}^{3}t_{i}\left[\cos(\bk\cdot\ba_{i})s_{0}\sigma_{x}
	+\sin(\bk\cdot\ba_{i})s_{0}\sigma_{y}\right]\nonumber\\
	&&+2\lambda_{\rm so}\sum_{i}\sin(\bk\cdot\bb_{i})s_{z}\sigma_{z}
	+(\bM\cdot\bs)\sigma_{z}.
	\label{KMH}
\end{eqnarray}
Above the first line describes the nearest-neighbor hoppings, the first term
in the second line represents the intrinsic spin-orbit coupling involving next-nearest-neighbor
hoppings, and the last term
denotes the exchange field associated with the antiferromagnetic order.
$(s_{0};s_{x},s_{y},s_{z})$ and $(\sigma_{0};\sigma_{x},\sigma_{y},\sigma_{z})$ are the identity matrix and Pauli matrices in the spin and
sublattice subspaces, respectively. The three nearest-neighbor lattice vectors are given by $\ba_{1}=a(0, 1)$, $\ba_{2}=\frac{a}{2}(\sqrt{3}, -1)$, $\ba_{3}=\frac{a}{2}(-\sqrt{3},-1)$, and the three next-nearest-neighbor lattice vectors are determined by $\ba_{i}$ through
the relation: $\bb_{1}=\ba_{2}-\ba_{3}$, $\bb_{2}=\ba_{3}-\ba_{1}$ and $\bb_{3}=\ba_{1}-\ba_{2}$.
For notational simplicity, we set the lattice constant $a=1$ throughout.

\begin{figure}[t]
	\centering
	\subfigure{
		\includegraphics[scale=0.4]{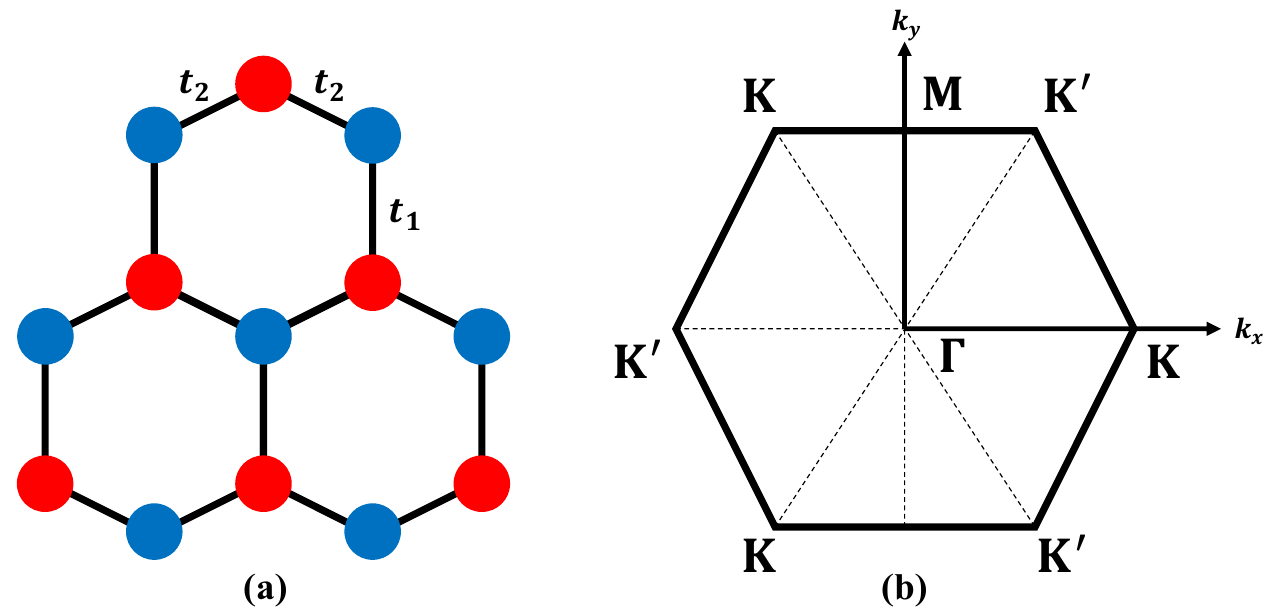}}
	\caption{(a) Schematic diagram of the honeycomb lattice with a specific type of lattice anisotropy that
 preserves the mirror symmetry about the $xz$ plane. Blue and red dots refer to $A$ and $B$ sublattices, respectively. (b) The Brillouin zone
 and some high-symmetric points.}
	\label{Fig1}
\end{figure}

Due to the antiferromagnetic exchange field, the above Hamiltonian does not have {\zb the spinful} time-reversal symmetry {\zb(symmetry 
operator is $\mathcal{T}=-is_{y}\sigma_{0}\mathcal{K}$ with $\mathcal{K}$
the complex conjugate operator)} and inversion symmetry {\zb ($\mathcal{P}=s_{0}\sigma_{x}$)}.
However, the Hamiltonian has their combination, the spinful $\mathcal{PT}$ symmetry. The symmetry operator
is $\mathcal{PT}=-is_{y}\sigma_{x}\mathcal{K}$, where $\mathcal{K}$ denotes
the complex conjugate operator and the symmetry operator satisfies $(\mathcal{PT})^{2}=-1$.
This combinational symmetry on one hand enforces
the band to be doubly degenerate, and on the other hand makes the Berry curvature identically vanishing. As a result, any Hall-type
effect with a Berry-curvature origin is expected to vanish.
When the \Neel vector is aligned in the $z$ direction and the hopping constants are isotropic,
the Hamiltonian also contains several important crystallographic symmetries that could have a strong
impact on the band topology and intrinsic NLHE, including the
$\mathcal{C}_{3z}$ rotation symmetry {\zb($\mathcal{C}_{3z}=e^{i\frac{\pi}{3}s_{z}\sigma_{0}}$)}, and the mirror symmetries about the $xy$ {\zb($\mathcal{M}_{z}=is_{z}\sigma_{0}$)} and $xz$ {\zb($\mathcal{M}_{y}=is_{y}\sigma_{x}$)} planes. To be general,
we incorporate lattice anisotropy that can be caused by
intrinsic lattice corrugation or extrinsic strain, and assume that the \Neel vector can point to any direction.
To be specific, for the lattice anisotropy, we set $t_{3}=t_{2}$,
but allow these two hopping constants to be different from $t_{1}$, as illustrated in
Fig.~\ref{Fig1}. For this type of lattice anisotropy, the $\mathcal{C}_{3z}$ rotation symmetry is broken once $t_{1}\neq t_{2}$, but
the mirror symmetry about the $xz$ plane remains.  For the convenience
of discussion, we introduce the ratio $\eta=t_{1}/t_{2}$ to characterize
the extent of lattice anisotropy. The more $\eta$ deviates from $1$, the
stronger the lattice anisotropy is.

\begin{figure*}[t]
	\centering
	\subfigure{
		\includegraphics[scale=0.5]{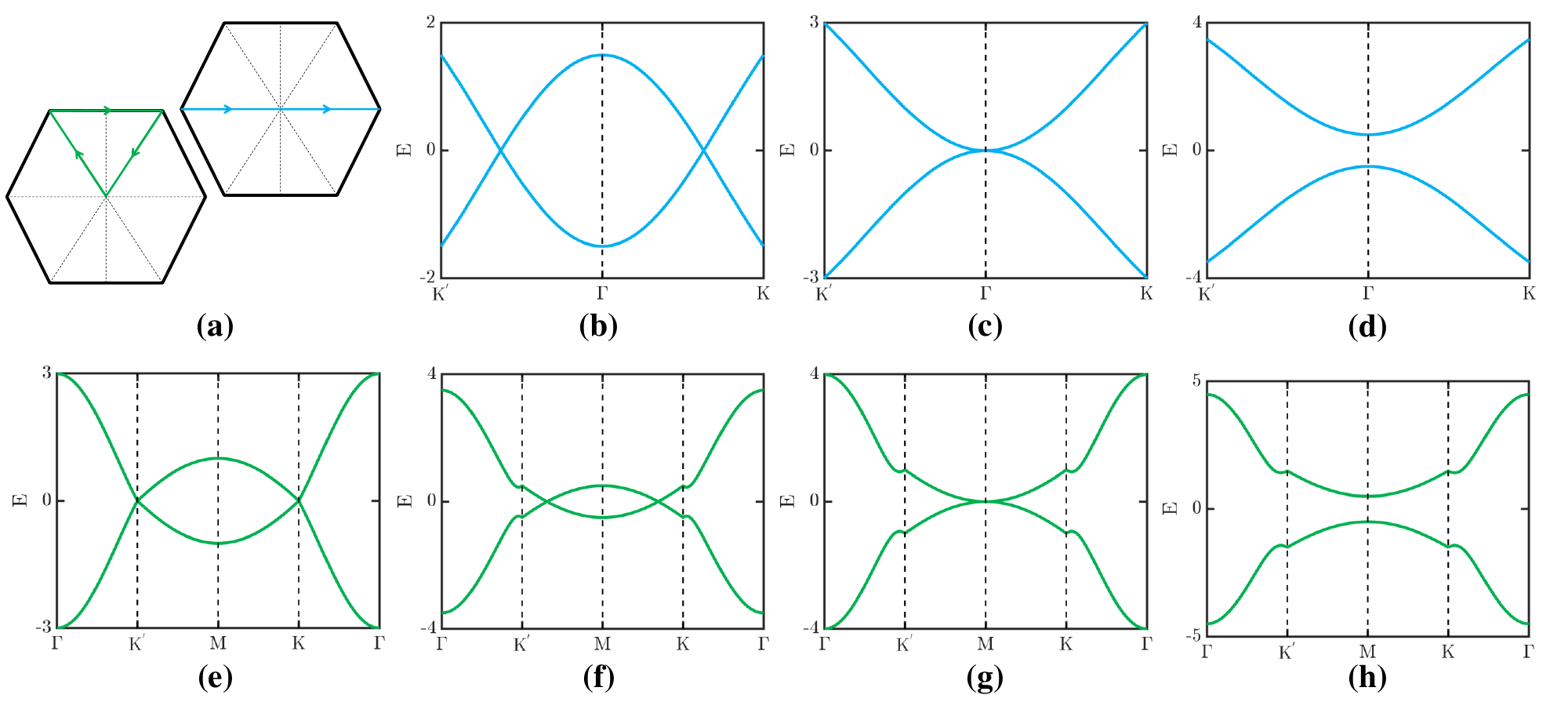}}
	\caption{The evolution of energy spectra along different paths in the Brillouin zone across the topological phase transition from a Dirac semimetal to an insulator. The light blue and green lines in (a) represent two different paths, with their corresponding energy spectra plotted in (b)-(d) and (e)-(h), respectively. When $\eta=1$, two Dirac cones are located at the two valleys $\mathbf{K}$ and $\mathbf{K}^{'}$. As $\eta$ deviates from $1$, the two Dirac points move away from the two valleys and eventually merge at the high symmetry point $\boldsymbol{\Gamma}$ ($\eta=-2$) or $\mathbf{M}$ ($\eta=2$), leading to the opening of the bulk energy gap once $|\eta|>2$. In (b)-(h), the values for $\eta$ are given as $-0.5$, $-2$, $-2.5$, $1$, $1.5$, $2$, and $2.5$. The shared parameters are given by $t_{2}=1$, $\lambda_{\rm so}=0$, and $M_{x}=M_{y}=M_{z}=0$.}
	\label{Fig2}
\end{figure*}

The band topology of the Hamiltonian in Eq.(\ref{KMH}) sensitively depends on the spin-orbit coupling
and antiferromagnetic exchange field. Without the spin-orbit coupling and antiferromagnetic exchange field,
it is known that the Hamiltonian realizes a Dirac semimetal
with two Dirac points at the two valleys, $\mathbf{K}$ and $\mathbf{K}'$, for
the case without lattice anisotropy~\cite{neto2009}. Weak lattice anisotropy shifts the locations of
the two Dirac points, but cannot annihilate them due to the protection of
a spinless $\mathcal{PT}$ symmetry (the corresponding symmetry operator is $\mathcal{PT}=s_{0}\sigma_{x}\mathcal{K}$,
satisfying $(\mathcal{PT})^{2}=1$). When
the lattice anisotropy reaches a critical condition ($\eta=\pm2$), the two
Dirac points converge and annihilate, and a further increase of the lattice anisotropy
opens an energy gap and drives the system to an insulator~\cite{Pereira2009graphene}, as illustrated in Fig.~\ref{Fig2}.

\begin{figure}[t]
	\centering
	\subfigure{
		\includegraphics[scale=0.34]{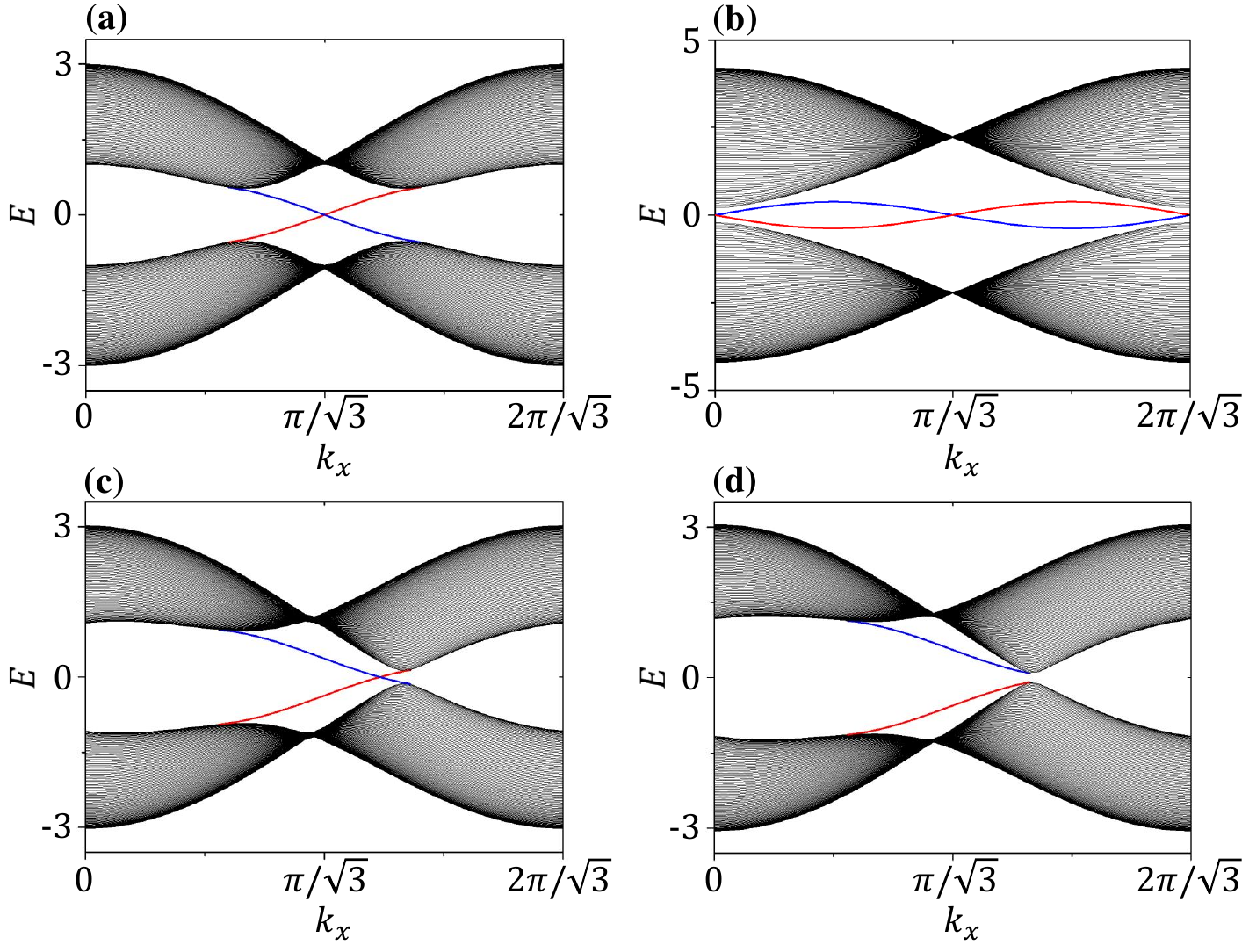}}
	\caption{(a) Quantum spin Hall insulator with {\zb a pair of }helical edge states. (b) Boundary-obstructed atomic insulator with boundary floating bands driven
by lattice anisotropy.
(c) The quantum spin Hall insulator remains stable when the antiferromagnetic exchange field is below
the critical value. (d) The quantum spin Hall insulator is transited to a trivial insulator when the antiferromagnetic exchange field
is beyond the critical value. For all figures, $t_{2}=1$, $\lambda_{\rm so}=0.1$, $M_{x}=M_{y}=0$ and the number of unit cells $N_{y}=100$. In (a), $\eta=1$ and $M_{z}=0$; In (b), $\eta=2.2$, and $M_{z}=0$; In (c), $\eta=1$ and $M_{z}=0.4$; In (d), $\eta=1$ and $M_{z}=0.6$. }
\label{Fig3}
\end{figure}

As long as the lattice anisotropy does not annihilate the two Dirac points, the presence of  spin-orbit coupling
will immediately gap out the Dirac points due to a lifting of the spinless $\mathcal{PT}$ symmetry, accompanying
with a direct transition from the Dirac semimetal to a quantum spin Hall insulator with helical edge states~\cite{Kane2005a,Kane2005b}, as
illustrated in Fig.~\ref{Fig3}(a).
Interestingly, recent works have shown that if the lattice anisotropy is strong, the quantum spin Hall insulator
does not become a featureless trivial insulator, instead,
it will evolve to a boundary-obstructed  atomic insulator which supports boundary floating bands or corner states for appropriate
geometry~\cite{Wang2021HOTI,Lahiri2023}, as illustrated in Fig.~\ref{Fig3}(b).

\begin{figure}[t]
	\centering
	\subfigure{
		\includegraphics[scale=0.5]{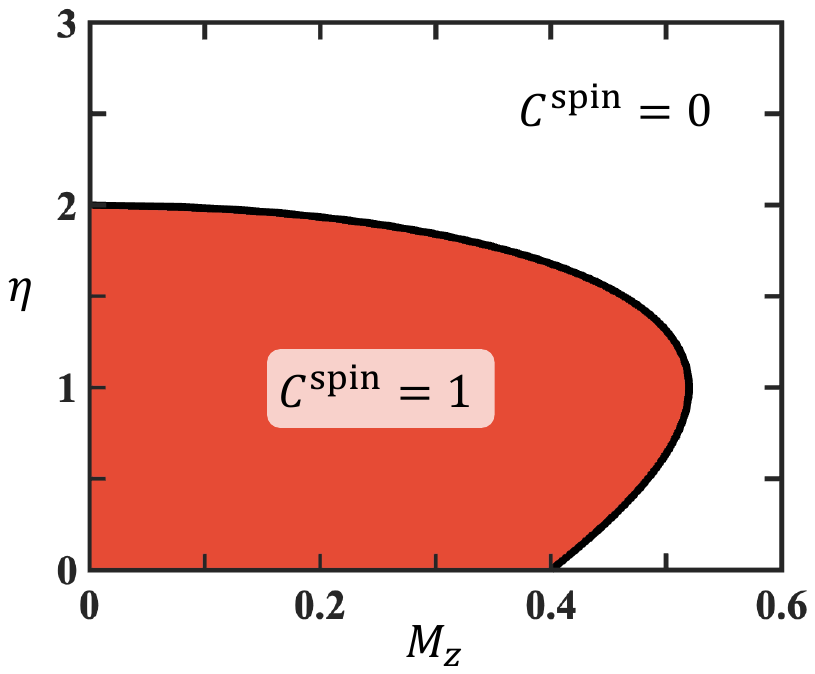}}
	\caption{The phase diagram with respect to $\eta$ and $M_{z}$. Parameters are $t_{2}=t_{3}=1$, $\lambda_{\rm so}=0.1$
and $M_{x}=M_{y}=0$. In the region with $C^{\text{spin}}=1$, the system is a quantum spin Hall insulator 
with a pair of helical edge states.}
\label{Fig4}
\end{figure}

In the quantum spin Hall regime, when the antiferromagnetic exchange field is also brought in, the time-reversal symmetry
protecting the helical edge states is broken. Nevertheless, the helical edge
states can remain stable if the \Neel vector is in the $z$ direction and
the strength of the exchange field is lower than a critical value ($M_{c}=3\sqrt{3}\lambda_{\rm so}$ for
the isotropic-hopping case), as illustrated in Figs.~\ref{Fig3}(c) and \ref{Fig3}(d).
There are two ways to understand the robustness of the helical edge states. The first one
is that the spin remains a good quantum number for this special case, therefore, the Hamiltonian
remains to be characterized by spin Chern number~\cite{Sheng2006QSHE}.  {\zb To be concrete, 
as $[s_{z},\mathcal{H}(\bk)]=0$, the Hamiltonian (\ref{KMH}) can be decomposed as the direct sum of two 
independent parts in accordance with the two eigenvalues of 
$s_{z}$, i.e., $\mathcal{H}(\bk)=\mathcal{H}_{s_{z}=1}(\bk)\oplus\mathcal{H}_{s_{z}=-1}(\bk)$,
where 
\begin{eqnarray}
\mathcal{H}_{s_{z}=s}(\bk)&=&\sum_{i=1}^{3}t_{i}\left[\cos(\bk\cdot\ba_{i})\sigma_{x}+\sin(\bk\cdot\ba_{i})\sigma_{y}\right]\nonumber\\
&&+s\left[2\lambda_{\rm so}\sum_{i=1}^{3}\sin(\bk\cdot\bb_{i})+M_{z}\right]\sigma_{z}\label{spinH}
\end{eqnarray}
with $s=\pm1$. The two-band Hamiltonians can be expressed as the form: $\bd^{s}(\bk)\cdot\boldsymbol{\sigma}$, with the three components 
of $\bd^{s}(\bk)$ being the coefficients before the corresponding Pauli matrices. For each spin sector, the Hamiltonian 
lacks time-reversal symmetry and is hence characterized by the Chern number,
\begin{eqnarray}
	C_{s}=\frac{1}{4\pi}\int_{\rm BZ}dk_{x} dk_{y}\frac{\bd^{s}(\bk)\cdot\left[\partial_{k_{x}}\bd^{s}(\bk)\times\partial_{k_{y}}\bd^{s}(\bk)\right]}{|\bd^{s}(\bk)|^{3}}.
\end{eqnarray}
As the two spin sectors are related by time-reversal symmetry, one has $C_{+1}=-C_{-1}$. While 
the total Chern number defined as $C_{T}=C_{+1}+C_{-1}$ is forced to be zero, the spin Chern number,
which is defined as $C^{\text{spin}}=(C_{+1}-C_{-1})/2$~\cite{Sheng2006QSHE}, can be nonzero. The spin Chern number 
counts the number of pairs of helical edge states. It will 
not change its nontrivial value unless the bulk energy gap closes
when the antiferromagnetic exchange field reaches its critical value, as 
illustrated by the phase diagram shown in Fig.\ref{Fig4}.}
The second one is that the the Hamiltonian has the mirror symmetry about
the $xy$ plane ({\zb recall} $\mathcal{M}_{z}=is_{z}\sigma_{0}$)
for this special case.  This mirror symmetry can also protect the helical edge
states as a mirror Chern number can be defined to characterize the Hamiltonian~\cite{teo2008}.
For this Hamiltonian, the mirror Chern number is just equal to
the spin Chern number, {\zb as the two different mirror-sector Hamiltonians 
obtained by decomposing the Hamiltonian in accordance with the two opposite eigenvalues of $\mathcal{M}_{z}$ are exactly the same as 
the two spin-sector Hamiltonians given in Eq.(\ref{spinH})}. In other words,
the topological gapped phase can be equivalently interpreted as either a quantum spin Hall insulator or a topological mirror insulator.

Once the direction of the \Neel vector deviates away from the $z$ direction, the $\mathcal{M}_{z}$ mirror symmetry
is broken and the spin is also no longer a conserved quantity. As a result, the helical edge
states will immediately be gapped due to the lack of any symmetry protection. Commonly,
the opening  of a gap to the edge states suggests that the resulting phase becomes a trivial insulator.
However, the trivialness is only strict in the first-order topology.
The resulting phase without gapless edge states is in fact not
completely topologically trivial.
In Ref.\cite{Miao2023}, the authors showed that, if the honeycomb lattice consists of two parts
with opposite in-plane \Neel vectors, despite the absence of gapless edge states in each part,
0D topological bound states will emerge at the ends of their interface.
All of these results reflect that the quantum geometric properties of honeycomb-lattice
materials have a sensitive dependence
on the interplay of lattice anisotropy, spin-orbit coupling
and antiferromagnetism.

\section{Intrinsic NLHE across topological phase transitions}\label{III}

As discussed above,
when the \Neel vector is aligned in $z$ direction, the system will undergo a topological phase transition
from a quantum spin Hall insulator to a trivial insulator with the increase of the antiferromagnetic
exchange field, or to a boundary-obstructed atomic insulator with the increase of lattice anisotropy.
In the following, we explore the behavior of the intrinsic NLHE across these two types of topological phase transitions.

Before start, we first give a brief review of the intrinsic
NLHE. In 2014, Gao, Yang and Niu showed that the electric field
can induce a first-order correction to the Berry curvature~\cite{Gao2014INHE}. As a result,
a second-harmonic Hall-type current can arise. The Hall-type current is
of the form $j_{\alpha}^{\rm int}=\chi_{\alpha\beta\gamma}^{\rm int}\mathcal{E}^{\beta}\mathcal{E}^{\gamma}$,
where $\mathcal{E}^{\beta}$ represents the electric-field component in the $\beta$ direction,
and $\chi_{\alpha\beta\gamma}^{\rm int}$ is a conductivity tensor independent
of scattering time.
The explicit expression of $\chi_{\alpha\beta\gamma}^{\rm int}$ is given by~\cite{Gao2014INHE,Liu2021INHE,Wang2021INHE}
\begin{eqnarray}
	\chi_{\alpha\beta\gamma}^{\rm int}=e^{3}\sum_{n}\int \frac{d^{D}k}{(2\pi)^{D}}\Lambda_{\alpha\beta\gamma}(\bk)\frac{\partial f(E_{n})}{\partial E_{n}},\label{tensor}
\end{eqnarray}
where $D$ is the dimension, $n$ is the band index, and $f(E_{n})$ is the equilibrium Fermi-Dirac distribution
function of the $n$th band. The tensor $\Lambda_{\alpha\beta\gamma}(\bk)$ is given by
\begin{eqnarray}	\Lambda_{\alpha\beta\gamma}^{(n)}(\bk)&=&v_{\alpha}^{(n)}(\bk)G_{\beta\gamma}^{(n)}(\bk)-v_{\beta}^{(n)}(\bk)G_{\alpha\gamma}^{(n)}(\bk),\label{AST}
\end{eqnarray}
where $v_{\alpha}^{(n)}=\partial E_{n}/\partial k_{\alpha}$ is the group velocity of the
$n$th band, and $G_{\beta\gamma}^{(n)}$ is of the form\cite{Gao2014INHE,Liu2021INHE,Wang2021INHE}
\begin{eqnarray}
	G_{\beta\gamma}^{(n)}(\bk)=2 {\rm Re} \sum_{m\neq n}\frac{A_{nm,\beta}(\bk)A_{mn,\gamma}(\bk)}{E_{n}(\bk)-E_{m}(\bk)}.
	\label{def: BCP}
\end{eqnarray}
Above $A_{nm,\beta}(\bk)=i\langle u_{n}(\bk)|\partial_{k_{\beta}}u_{m}(\bk)\rangle$ with $n\neq m$ is
the interband Berry connection. It is noteworthy that ${\rm Re} \sum_{m\neq n}A_{nm,\beta}(\bk)A_{mn,\gamma}(\bk)$
corresponds to the quantum metric of the $n$th band, suggesting the
quantum metric origin of this second-order response. In Eq.(\ref{tensor}), the derivative
of the Fermi-Dirac distribution function indicates that this effect is a Fermi-surface
property.
From Eq.(\ref{AST}),
it is easy to see that the tensor $\Lambda_{\alpha\beta\gamma}^{(n)}$, and so
the conductivity tensor $\chi_{\alpha\beta\gamma}^{\rm int}$,
is antisymmetric about the first two subscripts, i.e., $\chi_{\alpha\beta\gamma}^{\rm int}=-\chi_{\beta\alpha\gamma}^{\rm int}$,
suggesting that the resulting current is a Hall-like current. Because of this property,
the conductivity tensor only have two independent components in 2D, including
$\chi_{xyx}^{\rm int}$ and $\chi_{xyy}^{\rm int}$. On the other hand,
the energy difference between bands in the denominator of Eq.(\ref{def: BCP}) implies that
$G_{\beta\gamma}^{(n)}$ should be prominent near the band edge, and
a decrease in the band energy gap could benefit the enhancement of this effect.
Therefore, when the system is close to a topological phase transition, the intrinsic
NLHE is expected to be prominent.

Let us first focus on the topological phase transition driven by the $z$-directional antiferromagnetic exchange field.
To simplify the discussion, we consider the lattice anisotropy to be weak for this case. Accordingly,
the band edge will be located near one of the two valleys when the system is close to the topological
phase transition.
By an expansion of the bulk Hamiltonian around the two valleys and only keeping the leading-order terms,
we find that the corresponding low-energy Hamiltonians are given by
\begin{eqnarray}
	\mathcal{H}_{\chi}(\bq)&=&-\chi\frac{3t_{2}}{2}q_{x}\sigma_{x}+\left(t_{1}+\frac{t_{2}}{2}\right)q_{y}\sigma_{y}+M_{z}s_{z}\sigma_{z}\nonumber\\
	&&-3\sqrt{3}\chi\lambda_{\rm so}s_{z}\sigma_{z},\label{low}
\end{eqnarray}
where $\chi=1$ for the valley $\mathbf{K}$ and $-1$ for the valley $\mathbf{K}^{'}$. It is easy to see
that the energy gap gets closed at $\mathbf{K}$ if $M_{z}=M_{c}\equiv3\sqrt{3}\lambda_{\rm so}$, or
at $\mathbf{K}'$ if $M_{z}=-M_{c}$ (a more accurate analysis finds out that the critical value of the exchange field
is $M_{c}'=\lambda_{\rm so}(2+\eta)\sqrt{4-\eta^{2}}$).

\begin{figure}[t]
	\centering
	\subfigure{
		\includegraphics[scale=0.45]{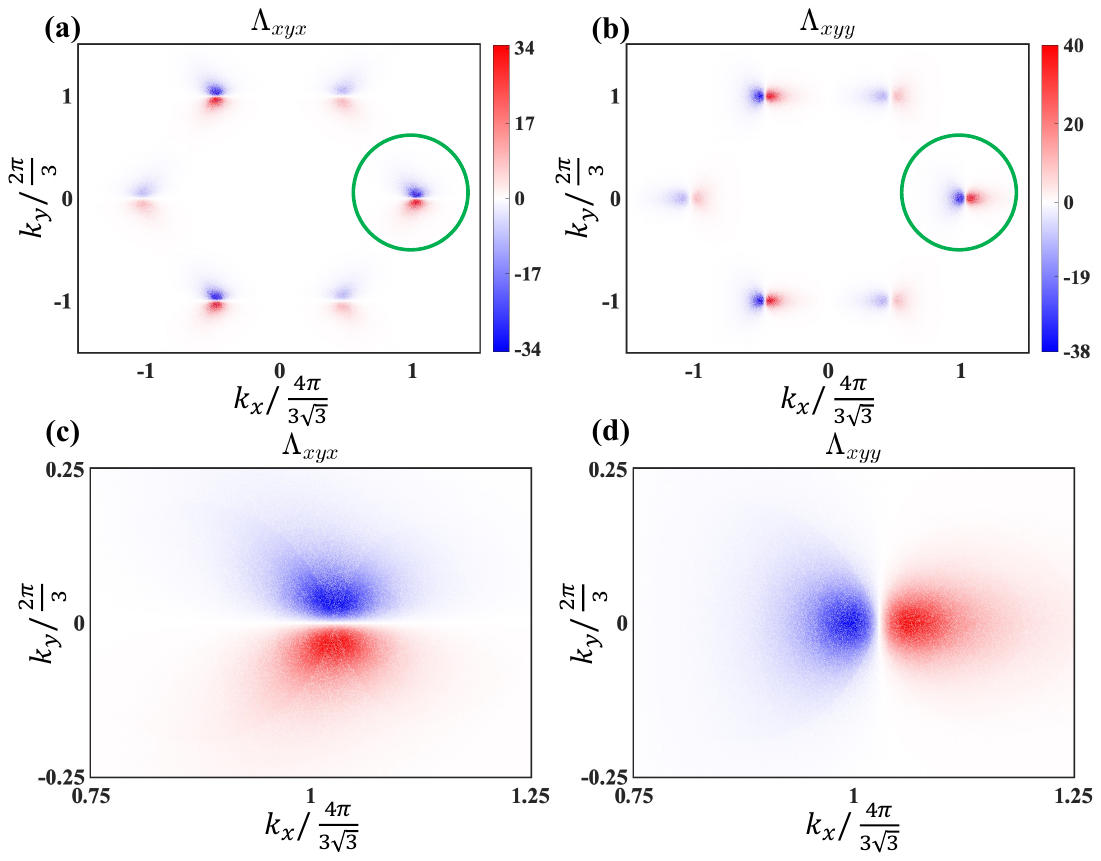}}
	\caption{The momentum-space distributions of the two geometric quantities, $\Lambda_{xyx}$ and $\Lambda_{xyy}$, for the conduction bands. (c) and (d) are the zoom-in plot of the area enclosed by the green circle in (a) and (b), respectively. The parameters are given by $t_{2}=1$, $\eta=1.1$, $\lambda_{\rm so}=0.05$, $M_{x}=M_{y}=0$ and $M_{z}=-0.2+\lambda_{\rm so}(2+\eta)\sqrt{4-\eta^{2}}$.}
	\label{Fig5}
\end{figure}

Although the low-energy Hamiltonians above can capture the topological phase transition, it has higher symmetry
than the full lattice Hamiltonian in Eq.(\ref{KMH}), and the symmetry emergent from
the leading-order approximation will force the intrinsic nonlinear Hall conductivity tensor (INLHCT)
to vanish identically. To correctly obtain the INLHCT, we {\zb always} adopt the full lattice Hamiltonian {\zb for calculations}.
It is noteworthy that if the lattice anisotropy is
absent, the $\mathcal{C}_{3z}$ rotation symmetry will force all components of the INLHCT to vanish identically even for
the full lattice Hamiltonian. When the lattice anisotropy is present,
the remaining mirror symmetry about the $xz$ plane forces the component $\chi_{xyx}^{\rm int}$
to vanish identically. Therefore, only the component $\chi_{xyy}^{\rm int}$ needs to be considered.
To intuitively see that $\chi_{xyx}^{\rm int}$ is forced to vanish while $\chi_{xyy}^{\rm int}$
is not, we plot the distribution of the geometric quantity $\Lambda_{xyx}(\bk)$ and
$\Lambda_{xyy}(\bk)$ in the Brillouin zone, as shown in Fig.~\ref{Fig5}. From Figs.~\ref{Fig5}(a) and \ref{Fig5}(c),
one sees that $\Lambda_{xyx}(\bk)$ is odd about $k_{y}$, leading to the vanishing of
$\chi_{xyx}^{\rm int}$ after the integration over the Fermi surface. In contrast,
Figs.~\ref{Fig5}(b) and \ref{Fig4}(d) show that $\Lambda_{xyy}(\bk)$ is even about $k_{y}$,
hence a nonzero $\Lambda_{xyy}(\bk)$ is permitted.

\begin{figure}[t]
	\centering
	\subfigure{
		\includegraphics[scale=0.33]{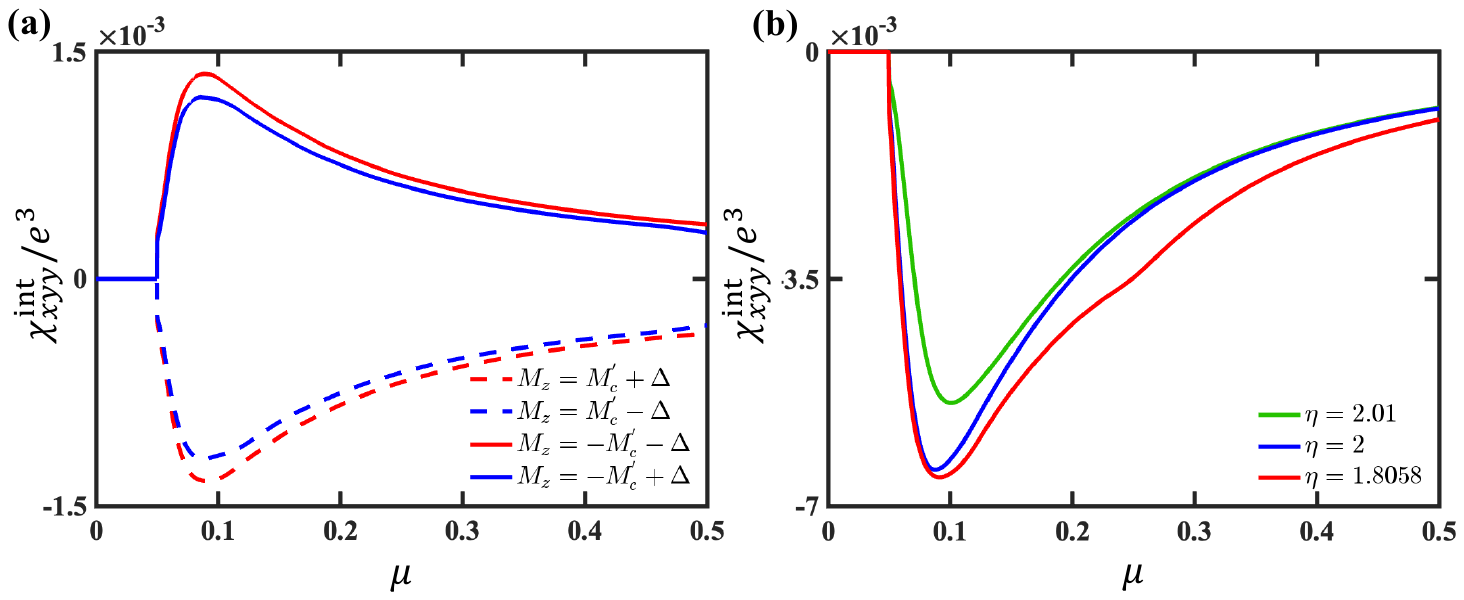}}
	\caption{The intrinsic NLHE before and after the topological phase transition from a quantum spin Hall insulator to a trivial insulator [(a)] and to a boundary-obstructed atomic insulator [(b)]. (a) The increase in the absolute value of $M_{z}$ across the critical value, $M_{c}'=\lambda_{\rm so}(2+\eta)\sqrt{4-\eta^{2}}$, renders the system topologically trivial. The parameters are given by $\eta=1.1$, $t_{2}=1$, $\lambda_{\rm so}=0.05$ and $M_{x}=M_{y}=0$. $\Delta=0.05$ refers
to half of the size of the bulk energy gap. (b) The quantum spin Hall insulator transits  into a boundary-obstructed atomic insulator as $\eta$ increases beyond $2$. In order to evaluate the impact of the lattice anisotropy on the intrinsic NLHE, we maintain the gap around $0.1$ by adjusting the strength of the lattice anisotropy. The parameters are given by $t_{2}=1$, $\lambda_{\rm so}=0.05$, $M_{x}=M_{y}=0$ and $M_{z}=0.11$.}
	\label{Fig6}
\end{figure}

The numerical results for the INLHCT across the topological phase transition are shown in Fig.~\ref{Fig6}.
Several prominent conclusions can be read from Fig.~\ref{Fig6}(a). First, the INLHCT does not change sign in the weakly-doped regime
when the system transits from a quantum spin Hall insulator to a trivial insulator. This is quite
different from the {\zb Berry-curvature-dipole} NLHE which shows a sign change when an inversion asymmetric
topological insulator transits to a trivial insulator~\cite{Du2018NHE,Sinha2022,Chakraborty2022}. 
The sign change of the {\zb Berry-curvature-dipole} NLHE
is simply due to the sign change of the Berry curvature across the
topological phase transition. However, for topological phase transitions characterized by
a Dirac Hamiltonian of the form  in Eq.(\ref{low}), it is easy to find that all components of
the quantum metric tensor do not change sign when the Dirac mass changes sign. As the intrinsic
NLHE is connected to the quantum metric, this explains why the INLHCT preserves
its sign across this class of topological phase transitions. In a previous work,
we have shown in the context of Hopf insulators that the INLHCT will change sign
across a topological phase transition  with the change of Hopf invariant~\cite{Zhuang2023NLHE}. Therein,
the critical point is different from a Dirac point in many aspects, and
it turns out that the quantum metric and Berry curvature are closely connected.
These results suggest that whether the INLHCT changes sign or not is not a universal property,
but depends on the type of the topological phase transition.
A second conclusion can be obtained from Fig.~\ref{Fig6}(a) is that the INLHCT switches sign
when the antiferromagnetic exchange field reverses its direction, which is expected
as the effect is time-reversal-odd. Last but not the least, {\zb under the condition of the same 
bulk energy gap, the INLHCT can be enhanced by increasing the antiferromagnetic exchange field to cross
the topological phase transition.}

Next we consider the topological phase transition from a quantum spin Hall insulator
to a boundary-obstructed atomic insulator driven by the lattice anisotropy. In Fig.~\ref{Fig2}, we have
shown that, in the absence of spin-orbit coupling and antiferromagnetic exchange field,
the two Dirac points will merge together at the $\mathbf{M}$ point when $\eta=2$ or  at
the $\boldsymbol{\Gamma}$ point when $\eta=-2$ and form a critical semi-Dirac point. Since $\eta=-2$ means
that the hopping constants  $t_{1}$ and $t_{2}$ take opposite signs, which is not very realistic
for a quantum material, below we will focus on the critical region at the neighborhood of $\eta=2$,
and explore the behavior of the INLHCT across the topological phase transition driven by lattice
anisotropy.

As the spin-orbit coupling vanishes at $\mathbf{M}$, a time-reversal invariant momentum, the
mergence of Dirac points at this point when $\eta=2$ indicates that the topological
phase transition is associated with the close of energy gap at $\mathbf{M}$.
Therefore, we can do a low-energy expansion of the lattice Hamiltonian around
this point. To capture the mergence of Dirac points, we keep the momentum up to the second order.
The low-energy Hamiltonian is given by
\begin{eqnarray}
	\mathcal{H}_{\mathbf{M}}(\bq)&=&t_{2}\left\{\left[\left(1-\frac{\eta}{2}\right)-\frac{3}{8}q_{x}^{2}\right]
-\frac{\sqrt{3}(1+\eta)}{2}q_{y}\right\}s_{0}\sigma_{x}\nonumber\\
&&-t_{2}\left\{\sqrt{3}\left[\left(1-\frac{\eta}{2}\right)-\frac{3}{8}q_{x}^{2}\right]
+\frac{(1+\eta)}{2}q_{y}\right\}s_{0}\sigma_{y}\nonumber\\
&&+(4\sqrt{3}\lambda_{\rm so}q_{x}+M_{z})s_{z}\sigma_{z}.
\end{eqnarray}
Without the last term, the energy spectrum is given by
\begin{eqnarray}
E_{\pm}(\bq)=\pm t_{2}\sqrt{4\left[\left(1-\frac{\eta}{2}\right)-\frac{3}{8}q_{x}^{2}\right]^{2}+(1+\eta)^{2}q_{y}^{2}},
\end{eqnarray}
which displays the characteristic feature of a semi-Dirac point when $\eta=2$, namely,
the energy spectrum is quadratic in one direction and linear in the other direction~\cite{Satyam2020BCD,Dietl2008DP,Banerjee2009DP}.
Because the quantum metric and the density of states, two factors
determining the conductivity tensor, are quite different between the Dirac point and the semi-Dirac point,
different features are expected to show up in the INLHCT when the system undergoes this
topological phase transition.

In Fig.~\ref{Fig6}(b), we show the INLHCT under different strength of lattice anisotropy and
fixed spin-orbit coupling and exchange field. Fixing the band gap by adjusting the strength of lattice anisotropy,
we find that the INLHCT still does not change sign across
the topological phase transition. However, the INLHCT is considerably enhanced near this
topological phase transition even for a weak exchange field.
The result suggests that the lattice anisotropy can be applied
as an effective factor to engineer strong intrinsic NLHE.

\begin{figure}[t]
	\centering
	\subfigure{
		\includegraphics[scale=0.35]{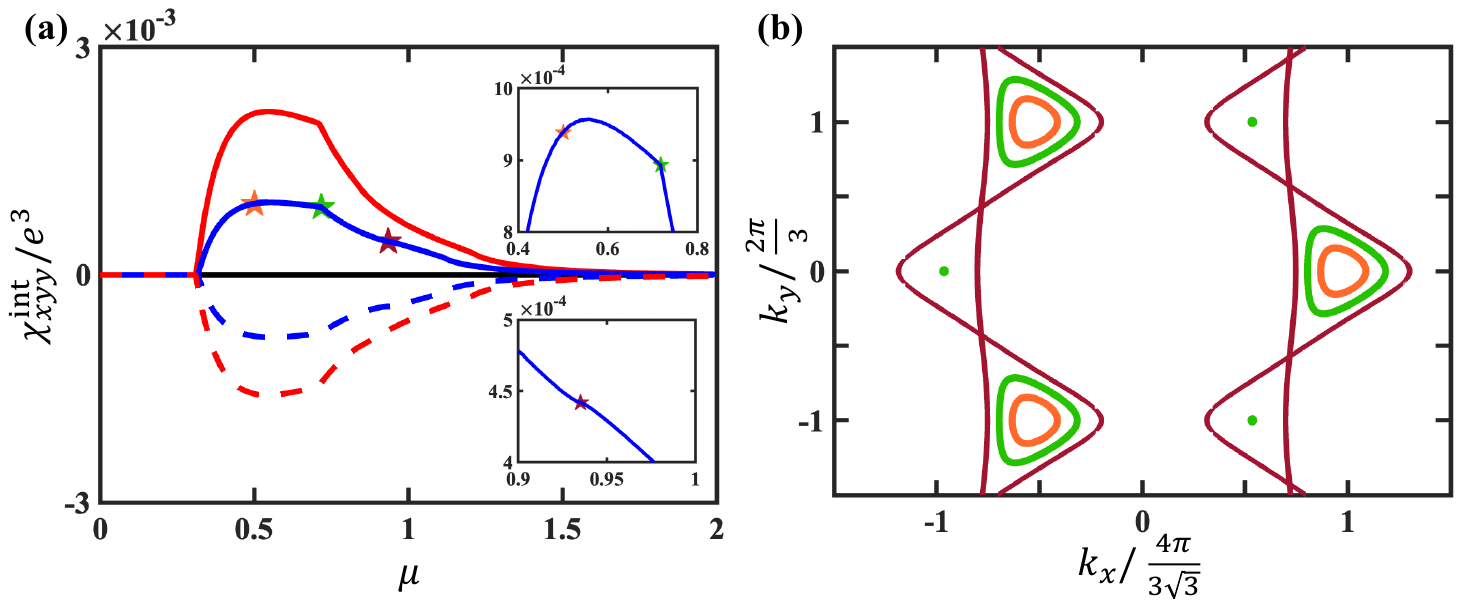}}
	\caption{ (a) $\chi_{xyy}^{\rm int}$ as a function of the chemical potential $\mu$; (b) The corresponding evolution of Fermi surface for the
blue solid line in (a).  In (a),  the red solid line, blue solid line, black solid line, blue dash line and red dash line refer to cases with $t_{2}=1$, $\eta=0.8$, $0.9$, $1$, $1.1$ and $1.2$, respectively. In (b), Fermi surfaces corresponding to $\mu=0.5$, $0.718$ and $0.935$ are plotted in orange, green and dark red, respectively. Shared parameters: $t_{2}=1$, $\lambda_{\rm so}=0.1$, $M_{x}=M_{y}=0$ and $M_{z}=0.2$.}
	\label{Fig7}
\end{figure}

\section{Detecting basic material properties via intrinsic NLHE}\label{IV}

\subsection{Manifestation of Lifshitz transitions}

The band structure of the honeycomb-lattice model is interesting
not only for its nontrivial topology, but also for properties like valley polarization and the existence of van Hove singularities
carrying divergent density of states~\cite{Hove1953}. By adjusting the Fermi level, both the valley polarization and
the van Hove singularities will manifest through the Lifshitz transitions, the change of Fermi surfaces in topology~\cite{Volovik2017}.

In Fig.~\ref{Fig7}, we show the evolution of INLHCT and Fermi surface with respect to the chemical potential on the left and right
panel, respectively. Three interesting features can be read from Fig.~\ref{Fig7}(a). The first one is that the INLHCT vanishes identically
when $\eta=1$, revealing that the breaking of the $\mathcal{C}_{3z}$ rotation symmetry is necessary for observing this effect.
The second one is the sign change of the INLHCT when $\eta$ goes across $1$, indicating that applying tensile or compressive strain
can tune both the magnitude and the direction of the nonlinear Hall current. The second one
is the existence of kinks on the INLHCT curves, as highlighted by the two stars
in green and dark red on the blue solid curve. By plotting the Fermi surfaces under
the chemical potential corresponding to the three stars of different colors,
we find that the kink highlighted by the green star corresponds to a Lifshitz transition
with a new Fermi surface emerging at the $\mathbf{K}'$ valley (the small green dots in Fig.~\ref{Fig7}(b)).
This result suggests that the intrinsic NLHE is valley-polarized for chemical potential below this value
(there is only one Fermi surface at the $\mathbf{K}$ valley, see the triangle-shaped orange ring in
Fig.~\ref{Fig7}(b)). For the kink
highlighted by the dark red star, it corresponds to a Lifshitz transition with the touching of the two Fermi surfaces
centered at $\mathbf{K}$ and $\mathbf{K}'$ valleys (see the Fermi surfaces in dark red). The touching point
is a saddle point, which corresponds to a van Hove singularity. The results above suggest that the intrinsic
NLHE, as a Fermi-surface property depending on the density of states, can detect the Lifshitz transitions
which is associated with a dramatic change of Fermi surface and the presence of  non-analyticity
in the density of states.

\begin{figure}[t]
	\centering
	\subfigure{
		\includegraphics[scale=0.5]{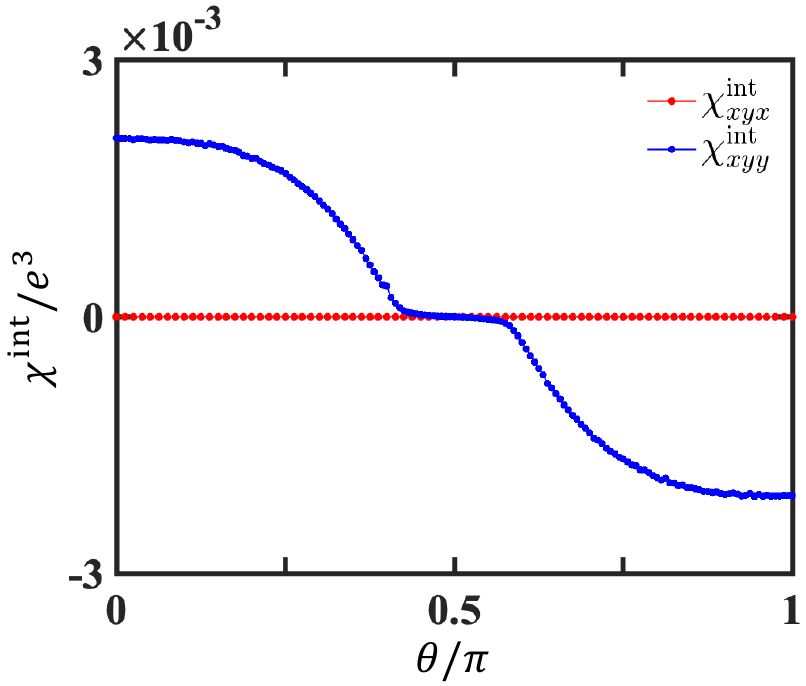}}
	\caption{$\chi_{xyy}^{\rm int}$ and $\chi_{xyx}^{\rm int}$ as a function of the polar angle $\theta$. The parameters are $\eta=0.8$, $t_{2}=1$, $\lambda_{\rm so}=0.1$, $\mu=0.5$ and the magnitude of the exchange field is fixed to $0.2$, i.e., $\sqrt{M_{x}^{2}+M_{y}^{2}+M_{z}^{2}}=0.2$.}
	\label{Fig8}
\end{figure}

\subsection{Detecting the \Neel vector}

Now we move to explore the dependence of INLHTC on the direction
of the \Neel vector. Without the spin-orbit coupling,
the Hamiltonian has spin-rotation symmetry and the intrinsic NLHE does not depend on the direction of the \Neel vector. In other words,
the intrinsic NLHE cannot reflect the \Neel vector if there is no spin-orbit coupling to break the spin rotation symmetry.
For the spin-orbit coupling considered, the full spin rotation symmetry  is broken down to a fixed-axis rotation symmetry, i.e.,
the system is invariant only when the spin is rotated about the $z$ axis. Therefore, if we view the \Neel vector in the spherical
coordinate system, the INLHTC is expected to depend on
the polar angle but not on the azimuthal angle of the \Neel vector.

By fixing the value of all parameters and only changing the direction of the \Neel vector, we calculate the evolution of the INLHTC with respect to the polar angle of the \Neel vector and present the numerical result in Fig.~\ref{Fig8}. The result clearly shows an angle dependence,
suggesting the capability of the intrinsic NLHE to detect the information of the \Neel vector. In Fig.~\ref{Fig8}, another notable feature is
that the INLHTC vanishes when the polar angle $\theta$ equals $\pi/2$, which corresponds to that the \Neel vector lies in the
$xy$ plane. The vanishing of INLHTC is due to the emergence of {\zb an effective} spinless time-reversal symmetry at this specific polar angle.
{\zb To be specific, at $\theta=\pi/2$, we find that there exists an operator of the form $\tilde{\mathcal{T}}=s_{x}\sigma_{0}\mathcal{K}$, satisfying
$\tilde{\mathcal{T}}\mathcal{H}(\bk)\mathcal{T}^{-1}=\mathcal{H}(-\bk)$ and $\tilde{\mathcal{T}}^{2}=1$.
Physically, this effective time-reversal symmetry is  a combinational symmetry
composed of the mirror operation about the $xy$ plane and the spinful time-reversal operation, 
i.e., $\tilde{\mathcal{T}}=\mathcal{M}_{z}\mathcal{T}$. While
the mirror symmetry $\mathcal{M}_{z}$ and the spinful time-reversal symmetry $\mathcal{T}$ are independently broken by the in-plane exchange field, their combination
remains intact for this special case.} As mentioned, the intrinsic NLHE is a time-reversal-odd
effect, the emergence of this time-reversal symmetry thereby forces it to vanish.

\section{Discussions and conclusions}\label{V}

We have explored the intrinsic NLHE in 2D
honeycomb antiferromagnets with $\mathcal{P}\mathcal{T}$ symmetry. As a class of systems
supporting a number of interesting topological phases, we investigated the behavior of the intrinsic
NLHE across two types of topological phase transitions. We found that, unlike the {\zb Berry-curvature-dipole} NLHE,
the intrinsic NLHE does not switch direction when the system undergoes a Dirac-type topological phase transition,
suggesting that the intrinsic NLHE cannot be applied to detect such topological phase transitions.
Nevertheless, the intrinsic NLHE could become prominent near these topological phase transitions, owing
to that the quantum metric is inversely proportional to the band gap.
We found that the lattice anisotropy breaking the $\mathcal{C}_{3z}$ rotation symmetry is not only
necessary for the presence of intrinsic NLHE, but also serves as an effective factor to tune
its magnitude and direction. As a Fermi-surface property,
we found that the intrinsic NLHE can manifest Lifshitz transitions. Furthermore,
we found that the existence of spin-orbit coupling to lift the spin-rotation
symmetry is crucial for the intrinsic NLHE to detect the \Neel vector. Our findings
show that the 2D honeycomb antiferromagnets could serve as fertile ground
to study the intrinsic NLHE.

\section*{Acknowledgments}

This work is supported by the National Natural Science Foundation of China (Grant No. 12174455),  the Natural Science Foundation of Guangdong Province (Grant No. 2021B1515020026) and
the Guangdong Basic and Applied Basic Research Foundation
(Grant No. 2023B1515040023).

\bibliography{dirac}

\end{document}